\documentclass[prl,twocolumn,noshowpacs,preprintnumbers,amsmath,amssymb]{revtex4}

\usepackage{graphicx}
\usepackage{dcolumn}
\usepackage{bm}
\usepackage{color}
\usepackage{soul}
\usepackage{multirow}
\usepackage {amsthm}

\begin{document}

 \preprint{APS/123-QED}
\title{Fluctuation-free Quantum Battery Charging Enabled by Bell-like Entangled and Separable Charger States}

\author{Longyan Gong  $^{1,2}$}
\thanks{Email address: lygong@njupt.edu.cn.}
\author{Jing Zhang $^{1,2}$}
\author{Yongtao Li $^{1,2}$}
\affiliation{
 $^{1}$College of Science, Nanjing University of Posts and Telecommunications, Nanjing, 210003, China\\
 $^{2}$Jiangsu Provincial Engineering Research Center of Low Dimensional Physics and New Energy, Nanjing, 210003, China\\
 }
\date{\today}
\begin{abstract}
The implication of entanglement in quantum battery charging remains poorly understood. In addition, intrinsic quantum randomness poses a fundamental challenge to charging uniformity across distinct trials. To address these issues, we study a three-qubit system within a cascaded collision frame, with two qubits acting as the charger and the third serving as the battery. After each collision, an adaptive measurement is performed on the charger, yielding four conditional post-measurement battery states with identical expected stored energy. For arbitrary fixed charger states, we derive analytical expressions for the recurrence relation of stored energy. We find that stored-energy fluctuations depend on whether the dynamics relies on the battery phase \(\gamma_n\). For two Bell-like families of entangled states, together with their computational basis states termed Bell-like separable states, the dynamics is \(\gamma_n\)-independent and fluctuations vanish, whereas for other entangled and separable states, fluctuations emerge. These insights establish a paradigm for engineering charger states to design uniform-charging quantum-battery architectures despite inherent quantum randomness.
\end{abstract}

\maketitle
\emph{Introduction}---Quantum batteries (QBs) store usable energy in controlled quantum states~\cite{CA24,FE26}. They supply power for quantum architectures, including quantum computers, quantum sensors, and quantum simulators~\cite{AU22,CA25}. Several proof-of-principle demonstrations have been realized, including spin systems~\cite{JO22}, superconducting qubits~\cite{YU24}, photonic systems~\cite{ZH23,LU25}, and microcavities~\cite{QU22,HY26}. QBs represent a promising energy technology~\cite{TA26,KH26}. They lie at the intersection of quantum thermodynamics, quantum information science, and quantum engineering~\cite{CA24,FE26,AU22}. Since Alicki and Fannes first introduced the concept in 2013, QBs have become one of the most active research topics in the field~\cite{AL13}.

Stored energy, charging power, charging precision, stability, and ergotropy are important figures of merit for assessing the performance of a quantum battery~\cite{CA24,FE26}. The roles of quantum resources, such as entanglement, correlations, and coherence, have attracted extensive attention. Of particular interest is dynamically generated entanglement: quantum states evolve under the dynamics of the charger-battery system, and entanglement characterizes these quantum states. The degree of such entanglement is time-varying and cannot be independently controlled. It can assist quantum charging~\cite{SH25}, yet it may be dispensable~\cite{KA20}, or even detrimental to quantum charging~\cite{KA20,LI21}. In fact, the charging performance is mainly governed by the interaction structure and collective effects, rather than the mere existence of entanglement~\cite{SH26}.

Recently, initial-state-input entanglement has drawn interest. Such entanglement can be prepared in the charger, intermediate mediators, or the battery. Specifically, initial entanglement between two photonic-cavity chargers can improve the stored energy and ergotropy under appropriate field intensities~\cite{AR22}. Entangled coupled-cavity arrays can serve as intermediate working mediators to boost charging performance~\cite{MA24}. Highly entangled initial states in batteries are not guaranteed to yield higher charging power, and detrimental entanglement may exist~\cite{GY24}. Meanwhile, initial coherence within chargers has likewise drawn interest~\cite{LI26}. Owing to the complicated many-body quantum dynamics, existing works are predominantly numerical, and key analytical insights remain lacking.

On the other hand, as practical devices, quantum batteries require high charging precision~\cite{FH18,CR20}, which is quantified by the energy variance of the final battery state. Calculated from the density matrix, this quantity encompasses both classical and quantum fluctuations. However, this ensemble variance does not fully characterize trial-to-trial fluctuations in individual charging realizations. Recently, Mitchison \emph{et al.} addressed this issue via stochastic master-equation simulations of quantum trajectories, where each trajectory represents a single realization of the charging process. They sought to mitigate stored-energy fluctuations across distinct trajectories using both instantaneous and delayed linear feedback; nevertheless, such fluctuations remain substantial~\cite{MI21}.

Very recently, within cascaded collision models, we proposed an adaptive-angle measurement on a single-qubit charger~\cite{GO26}. In this Letter, we consider a two-qubit charger coupled to a single-qubit QB. The two-qubit charger can be prepared in maximally entangled, partially entangled, or separable states. For an arbitrary fixed charger state, we derive an analytical expression for the stored battery energy. Using this expression, we investigate the influences of entanglement, coherence, relative phases, and populations of charger states throughout the charging process. Despite consisting of only two qubits, this minimal setup captures all essential features of more complex charging architectures.

\emph{Model}---The Hamiltonian of the charger-battery system reads
\begin{equation}
H=H_a+H_b+H_{ab},
\label{EQ_H3}
\end{equation}
where
\[
\begin{aligned}
&H_a=\omega_{a_1} n_{a_1}+\omega_{a_2} n_{a_2}, \quad H_b=\omega_b n_b,\\
&H_{ab}=g\left(\sigma_{a_1}\sigma^\dagger_b+\sigma^\dagger_{a_1}\sigma_b\right)
+g\left(\sigma_{a_2}\sigma^\dagger_b+\sigma^\dagger_{a_2}\sigma_b\right).
\end{aligned}
\]
Here, $a_1$ and $a_2$ label the two qubits of charger $a$, and $b$ labels the battery. The number operator is defined as $n_m=\sigma^\dagger_m\sigma_m$, with $\sigma_m=|0\rangle_m\langle 1|$, $\sigma^\dagger_m=|1\rangle_m\langle 0|$. The parameter $g$ denotes the coupling strength between the charger and the battery. For simplicity, we set $\omega_{a_1}=\omega_{a_2}=\omega_0$, with $\omega_b=1$ serving as the energy unit.

\emph{Collision, i.e., unitary evolution with interaction}---Each realization of the charging process begins with pure charger and battery states, denoted $|\phi_{a}(0)\rangle$ and $|\phi_b(0)\rangle$. The initial joint state reads
$|\Phi(0)\rangle = \sum_{i=0}^{7} \phi_i |i\rangle$, and time evolution follows $|\Phi(t)\rangle=e^{-iHt}|\Phi(0)\rangle$, taking the form
$|\Phi(t)\rangle = \sum_{i=0}^{7} \phi_i(t) |i\rangle$. This finite-time unitary evolution generated by $H$ constitutes one discrete collision between the charger and battery~\cite{CI22}.

\emph{Projective measurement and conditional states}---For the qubits \(a_1\) and \(a_2\), we define \(\mathrm{SU}(2)\) rotations
\begin{equation}
\begin{aligned}
U_1 &= \begin{pmatrix} \cos\frac{\xi_1}{2} & -e^{-i\eta_1}\sin\frac{\xi_1}{2}\\ e^{i\eta_1}\sin\frac{\xi_1}{2} & \cos\frac{\xi_1}{2} \end{pmatrix},\\
U_2 &= \begin{pmatrix} \cos\frac{\xi_2}{2} & -e^{-i\eta_2}\sin\frac{\xi_2}{2}\\ e^{i\eta_2}\sin\frac{\xi_2}{2} & \cos\frac{\xi_2}{2} \end{pmatrix},
\end{aligned}
\label{EQ_U12}
\end{equation}
where \(\xi_1,\xi_2\in[0,\pi]\) and \(\eta_1,\eta_2\in[0,2\pi)\). Taking the phase $\eta_1$ of $U_1$ as a reference, we set $\eta_1=0$.
Let $U_{12}=U_1 \otimes U_2$. We then introduce a projective basis
\begin{equation}
\big(|b_0\rangle,|b_1\rangle,|b_2\rangle,|b_3\rangle\big)^{\!\top} =U_{12}^{\!\top} \,\big(|00\rangle,|01\rangle,|10\rangle,|11\rangle\big)^{\!\top},
\label{EQ_4B}
\end{equation}
where the superscript \(\top\) denotes the transpose operation. The basis is determined by the parameters \(\xi_1, \xi_2, \eta_2\).

After the collision lasting for \(\tau\), i.e., at \(t=\tau\), the joint state is rewritten as
\[
|\Phi(\tau)\rangle=\sum_{m=0}^{3}|b_m\rangle\otimes\big(\alpha_m|0\rangle+\beta_m|1\rangle\big).
\]
Moreover,
\begin{equation}
\boldsymbol{\alpha}=U_{12}^{\dagger}\boldsymbol{\Phi}_0,\qquad \boldsymbol{\beta}=U_{12}^{\dagger}\boldsymbol{\Phi}_1,
\label{EQ_AB}
\end{equation}
where
\[
\begin{aligned}
&\boldsymbol{\alpha}=(\alpha_0,\alpha_1,\alpha_2,\alpha_3)^{\!\top}, \boldsymbol{\Phi}_0=(\phi_0(\tau),\phi_2(\tau),\phi_4(\tau),\phi_6(\tau))^{\!\top},\\
&\boldsymbol{\beta}=(\beta_0,\beta_1,\beta_2,\beta_3)^{\!\top}, \boldsymbol{\Phi}_1=(\phi_1(\tau),\phi_3(\tau),\phi_5(\tau),\phi_7(\tau))^{\!\top}.
\end{aligned}
\]
The probability of obtaining \(|b_m\rangle\) upon projective measurement of the charger is
\(\rho_m = |\alpha_m|^2 + |\beta_m|^2\). Then the conditional state of the battery is
\begin{equation}
|\phi^m_b(\tau)\rangle=(\alpha_m|0\rangle+\beta_m|1\rangle)/\sqrt{\rho_m}
\label{EQ_CS}
\end{equation}
for \(m=0,1,2,3\).

\emph{Adaptive measurements on chargers}---For the \(m\)-th branch, the stored energy of the battery is
\[
E_m=\langle\phi^m_b(\tau)|H_\text{b}|\phi^m_b(\tau)\rangle=|\beta_m|^2/\rho_m.
\]

In Ref.~\cite{GO26}, we developed an adaptive-measurement protocol for the charger using adaptive angles. Similarly, using Eqs.~\eqref{EQ_U12} and~\eqref{EQ_4B}, we may select suitable parameters $\xi_1,\xi_2,\eta_2$ to set the adaptive angles and construct a projective basis \(\{|b_m\rangle\}\). This basis ensures equal stored energies among the four branches, i.e.,
\begin{equation}
E^b_0=E^b_1=E^b_2=E^b_3.
\label{EQ_4E}
\end{equation}
This implies
\begin{equation}
\frac{|\beta_0|^2}{|\alpha_0|^2}=\frac{|\beta_1|^2}{|\alpha_1|^2}=\frac{|\beta_2|^2}{|\alpha_2|^2}=\frac{|\beta_3|^2}{|\alpha_3|^2}
=\frac{P_1(\tau)}{P_0(\tau)},
\label{EQ_4C}
\end{equation}
assuming \(|\alpha_m|^2\neq 0\), where \(P_0(\tau)=\sum_{m=0}^3|\alpha_m|^2\) and \(P_1(\tau)=\sum_{m=0}^3|\beta_m|^2\). According to Eq.~\eqref{EQ_AB},
\begin{equation}
P_1(\tau)=|\phi_1(\tau)|^2 + |\phi_3(\tau)|^2 + |\phi_5(\tau)|^2 + |\phi_7(\tau)|^2.
\label{EQ_P1}
\end{equation}
The conditional state in Eq.~\eqref{EQ_CS} can then be rewritten as
\begin{equation}
|\phi^m_b(\tau)\rangle=\sqrt{1-P_1(\tau)}|0\rangle+\sqrt{P_1(\tau)}e^{i \gamma_m}|1\rangle.
\label{EQ_CSN}
\end{equation}
These four states correspond to the four measurement branches indexed by $m$. An adaptive projective measurement on the charger collapses the joint state into one of them. The resulting post-measurement battery state serves as the initial state for the subsequent collision in the cascaded protocol.

For the first collision step, the stored energy of the battery reads \(E_{n=1}=P_1(\tau)\). Generalizing to arbitrary \(n\),
\begin{equation}
E_n=P_1(n\tau).
\label{EQ_E}
\end{equation}
Henceforth, we omit the argument \(n\tau\) of \(P_0(n\tau)\) and \(P_1(n\tau)\) unless needed.

\noindent\textbf{Proposition.}
Suppose \(P_0>0\) and \(P_1>0\). There exists \((\xi_1,\xi_2,\eta_2)\in\mathbb R^3\) such that
\[
|\beta_m|^2=\frac{P_1}{P_0}\,|\alpha_m|^2\qquad\forall\,m\in\{0,1,2,3\}.
\]

\noindent\textit{Proof.}
If \(\beta_m=\sqrt{\frac{P_1}{P_0}}e^{i\Gamma_m}\alpha_m\), then \(\boldsymbol{\beta}=\sqrt{\frac{P_1}{P_0}}\,T\boldsymbol{\alpha}\), where \(T=\mathrm{diag}(e^{i\Gamma_0},e^{i\Gamma_1},e^{i\Gamma_2},e^{i\Gamma_3})\). Using Eq.~\eqref{EQ_AB}, we obtain
\[
\boldsymbol{\Phi}_1=\sqrt{\frac{P_1}{P_0}}\,U_{12} T U_{12}^\dagger \boldsymbol{\Phi}_0.
\]
Define \(\boldsymbol{\Phi}'_0=\frac{\boldsymbol{\Phi}_0}{\sqrt{P_0}}\) and
\(\boldsymbol{\Phi}'_1=\frac{\boldsymbol{\Phi}_1}{\sqrt{P_1}}\),
which are normalized vectors. This yields the relation
\[
\boldsymbol{\Phi}'_1=U_{12} T U_{12}^\dagger \boldsymbol{\Phi}'_0,
\]
with \(U_{12}=U_1 \otimes U_2\).
Since \(U_1\) and \(U_2\) are arbitrary \(\mathrm{SU}(2)\) matrices parameterized by \((\xi_1,\xi_2,\eta_2)\) [see Eq.~\eqref{EQ_U12}], such parameters always exist so that \(U_{12}TU_{12}^\dagger\) transforms \(\boldsymbol{\Phi}'_0\) into \(\boldsymbol{\Phi}'_1\). This completes the proof.

\emph{Stored energies in the battery}---For cascaded collision models~\cite{CI22}, the battery undergoes successive collisions with a sequence of ``fresh'' chargers. Each charger is prepared in an identical normalized state $|\phi_a\rangle$, which is generally kept fixed, and each collision evolves over a time $\tau$.  Each charger state reads
\begin{equation}
\begin{aligned}
|\phi_a\rangle&=\sqrt{e_{00}}|00\rangle+\sqrt{e_{01}}\,e^{i\gamma_{01}}|01\rangle\\
&+\sqrt{e_{10}}\,e^{i\gamma_{10}}|10\rangle+\sqrt{e_{11}}\,e^{i\gamma_{11}}|11\rangle.
\end{aligned}
\label{EQ_phia}
\end{equation}
At the beginning of the \((n+1)\)-th collision, the battery in the state
\begin{equation}
|\phi_b(n\tau)\rangle=\sqrt{1-e_{n}}|0\rangle+\sqrt{e_{n}}\,e^{i\gamma_{n}}|1\rangle,
\label{EQ_phib}
\end{equation}
where
\begin{equation}
e_n = P_1(n\tau)
\label{EQ_en}
\end{equation}
and \(\gamma_n=\gamma_m\) for the measurement branch \(m\) realized right after the $n$-th collision, see Eq.~\eqref{EQ_CSN}.
The initial joint state reads
\(|\Phi(n\tau)\rangle=|\phi_a\rangle\otimes|\phi_b(n\tau)\rangle\).
In the \((n+1)\)-th collision, lasting for a duration \(\tau\), the joint state evolves as
\(|\Phi\big((n+1)\tau\big)\rangle=e^{-iH\tau}|\Phi(n\tau)\rangle\).
Subsequently, performing the adaptive measurement with angles \(\xi_1,\xi_2,\eta_2\) on the charger yields the post-measurement battery state
\begin{equation}
|\phi_b\big((n+1)\tau\big)\rangle=\sqrt{1-e_{n+1}}|0\rangle+\sqrt{e_{n+1}}\,e^{i\gamma_{n+1}}|1\rangle.
\label{EQ_phib1}
\end{equation}
According to Eq.~\eqref{EQ_en}, \(e_{n+1}=P_1\big((n+1)\tau\big)\). Thus
\begin{widetext}
\begin{equation}
\begin{aligned}
e_{n+1}
={}& e_n+\frac{4g^{2}}{\Omega^{2}}S_\tau^2\Bigg[
\big(e_{01}+e_{10}+2e_{11}-2e_n\big)
+\sqrt{e_{01}e_{10}}\,(1-2e_n)\cos(\gamma_{01}-\gamma_{10})
\Bigg] \\
&+\frac{4g}{\Omega}S_\tau\sqrt{e_n(1-e_n)}
\Bigg\{
\sqrt{e_{00} e_{01}}
\left[\frac{\Delta}{\Omega}S_\tau\cos\left(\gamma_n-\gamma_{01}\right)-C_\tau\sin\left(\gamma_n-\gamma_{01}\right)\right] \\
&\quad+\sqrt{e_{00} e_{10}}
\left[\frac{\Delta}{\Omega}S_\tau\cos\left(\gamma_n-\gamma_{10}\right)-C_\tau\sin\left(\gamma_n-\gamma_{10}\right)\right] \\
&\quad+\sqrt{e_{01}e_{11}}
\left[\frac{\Delta}{\Omega}S_\tau\cos\left(\gamma_n+\gamma_{01}-\gamma_{11}\right)-C_\tau\sin\left(\gamma_n+\gamma_{01}-\gamma_{11}\right)\right] \\
&\quad+\sqrt{e_{10}e_{11}}
\left[\frac{\Delta}{\Omega}S_\tau\cos\left(\gamma_n+\gamma_{10}-\gamma_{11}\right)-C_\tau\sin\left(\gamma_n+\gamma_{10}-\gamma_{11}\right)\right]
\Bigg\}.
\end{aligned}
\label{EQ_en1}
\end{equation}
\end{widetext}
Here, \(S_\tau = \sin\frac{\Omega\tau}{2}\) and \(C_\tau = \cos\frac{\Omega\tau}{2}\); \(\Delta=\omega_b-\omega_0\) is the detuning and \(\Omega=\sqrt{\Delta^2+8g^2}\) is the effective Rabi frequency. The stored energy of the battery reads
\begin{equation}
E_{n+1}=P_1\big((n+1)\tau\big)=e_{n+1}.
\label{EQ_Estore}
\end{equation}

\emph{Results and discussion}---In simulations, $10^5$ stochastic quantum trajectories, i.e., charging realizations, are sampled. Additional trajectories yield consistent averaged quantities.

For convenience of analysis, we initialize the battery as $|\phi_b(0)\rangle=|0\rangle$, i.e., $e_0=0$. We also fix $\Omega\tau/2=\pi/2$, which gives $S_\tau = 1$ and $C_\tau = 0$. As a benchmark example, each charger is prepared in $|\phi_a\rangle=e^{i\gamma_{11}}|11\rangle$, i.e., $e_{11}=1$. The detuning $\Delta=0$, so $\Omega^2=8g^2$. Equation~\eqref{EQ_en1} gives $e_{1}=1$, meaning full charging of the battery in a single collision step. This case recovers the standard Rabi model. In this Letter, we focus on off-resonant cases.

\emph{Entangled states}---For charger states given in Eq.~\eqref{EQ_phia}, the concurrence, a measure of two-qubit entanglement~\cite{HI97,WO98}, reads
\begin{equation}
C=2\left|\sqrt{e_{00}e_{11}}\,e^{i\gamma_{11}}-\sqrt{e_{01}e_{10}}\,e^{i(\gamma_{01}+\gamma_{10})}\right|.
\label{EQ_CC}
\end{equation}

(i) Bell-like-I entangled states. The charger state is
\begin{equation}
|\phi_a\rangle=\sqrt{e_{00}}|00\rangle+\sqrt{e_{11}}\,e^{i\gamma_{11}}|11\rangle.
\label{EQ_BS1}
\end{equation}
Its concurrence reads \(C=2\sqrt{e_{11}(1-e_{11})}\). From Eq.~\eqref{EQ_en1}, we obtain
\begin{equation}
e_n=e_{11}\left[1-\left(1-\frac{8g^{2}}{\Delta^{2}+8g^{2}}\right)^n\right],\quad E_n=e_n.
\label{EQ_Edef_BS1}
\end{equation}

\begin{figure}[!htbp]
\centering
 \includegraphics[width=0.5\columnwidth]{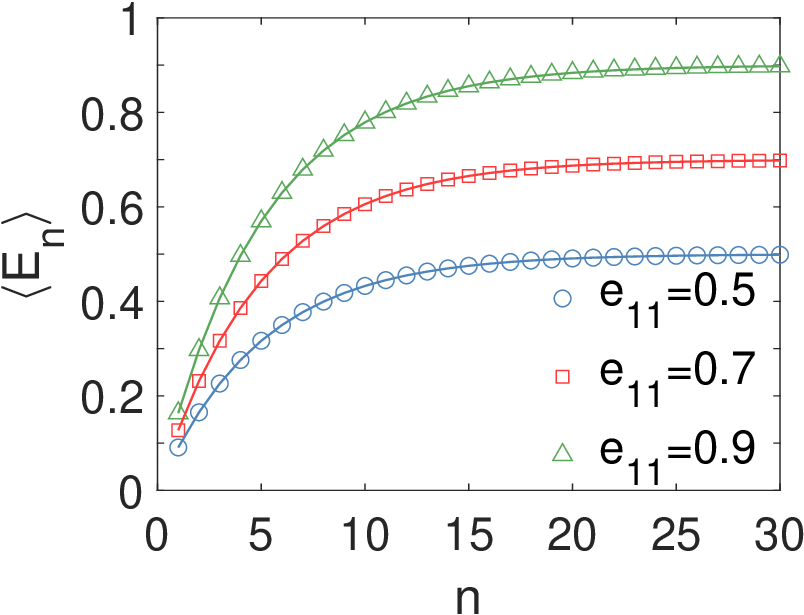}
\caption{Trajectory-averaged stored energy \(\langle E_n \rangle\) versus collision steps \(n\). Lines denote theoretical
predictions from Eq.~\eqref{EQ_Edef_BS1}. Parameters: $\Delta=0.7$, $g=0.05$, and $\gamma_{11}=\pi/7$.}
\label{Fig1}
\end{figure}

This deterministic expression for \(E_n\) indicates that stored-energy fluctuations \( \sigma(E_n) \) are fully suppressed, where \( \sigma(E_n)=\sqrt{\langle E_n^2\rangle-\langle E_n\rangle^2} \) and \( \langle X\rangle \) denotes the average of \( X \) over distinct trajectories. The charging rate depends solely on the system parameters \( \Delta \) and \( g \), and increasing \( \Delta \) slows down the charging process. Fig.~\ref{Fig1} shows that numerical simulations are in complete quantitative agreement with theoretical predictions. Notably, we obtain \(\sigma(E_n) < 5.0\times 10^{-14}\), consistent with vanishing fluctuations up to numerical precision (omitted from Fig.~\ref{Fig1}).

The stored-energy upper bound \(E^* = e_{11}\), so the concurrence satisfies \(C=2\sqrt{E^*(1-E^*)}\), i.e., \(E^*=\frac12\big(1\pm\sqrt{1-C^2}\big)\). For the maximally entangled state with \(C=1\), \(E^* = 1/2\). \(C\) increases with \(E^*\) for \(E^*<1/2\) and decreases for \(E^*>1/2\). Therefore, larger entanglement does not necessarily yield a higher saturation stored energy.

(ii) Bell-like-II entangled states. The charger state is
\begin{equation}
|\phi_a\rangle=\sqrt{e_{01}}\,e^{i\gamma_{01}}|01\rangle+\sqrt{e_{10}}\,e^{i\gamma_{10}}|10\rangle.
\label{EQ_BS2}
\end{equation}
Its concurrence reads \(C=2\sqrt{e_{01}e_{10}}\). From Eq.~\eqref{EQ_en1}, we obtain
\begin{equation}
e_n=\frac12\left[1-\left(1-\frac{8g^{2}}{\Delta^2+8g^2}Z\right)^n\right],\quad E_n=e_n,
\label{EQ_Edef_BS2}
\end{equation}
where \(Z \equiv 1+\sqrt{e_{01}e_{10}}\cos(\gamma_{01}-\gamma_{10})\). We define \(A=1-\frac{8g^{2}}{\Delta^2+8g^2}Z\), and \(A\) lies in \(\big[1-\frac{12g^2}{\Delta^2+8g^2},1-\frac{4g^2}{\Delta^2+8g^2}\big]\). The condition \(|A|<1\) always holds for nonvanishing \(g\), ensuring the sequence converges.

\begin{figure}[!htbp]
\centering
 \includegraphics[width=0.5\columnwidth]{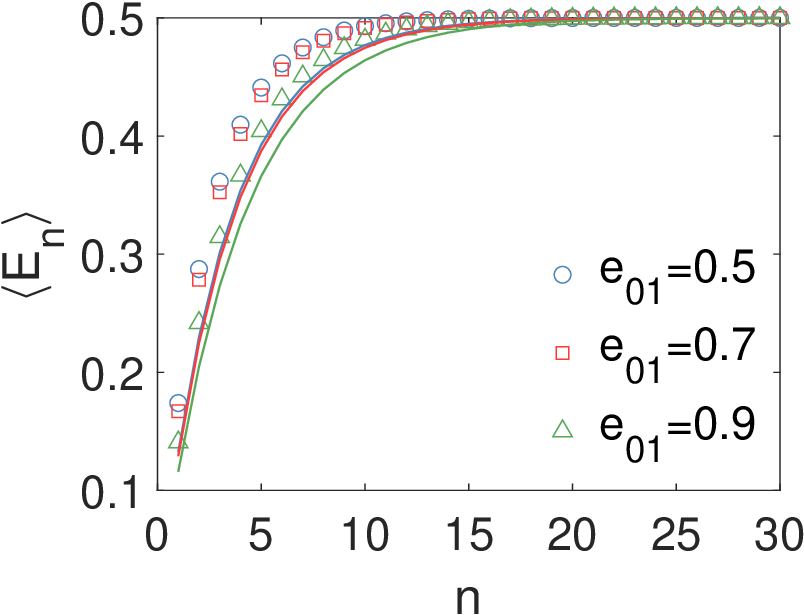}
\caption{Trajectory-averaged stored energy \(\langle E_n \rangle\) versus collision steps \(n\). Lines denote theoretical predictions from Eq.~\eqref{EQ_Edef_BS2}. Parameters: $\Delta=0.7$, $g=0.05$, \(\gamma_{01}=\pi/3\) and \(\gamma_{10}=\pi/5\).
} \label{Fig2}
\end{figure}

As the expression for \(E_n\) is deterministic, stored-energy fluctuations \(\sigma(E_n)\) are fully suppressed. The stored-energy upper bound is fixed at \(E^*=1/2\). As reflected in factor \(A\), the charging rate depends on \(\delta\gamma=\gamma_{01}-\gamma_{10}\), \(e_{01}\), and \(e_{10}\). Further, in terms of concurrence \(C\), \(Z=1+\tfrac12 C\cos(\delta\gamma)\), which indicates \(C\) can influence the charging rate but not \(E^*\). Simulation results in Fig.~\ref{Fig2} confirm the theoretical upper bound \(E^*=1/2\), though the trajectory-averaged stored energy lies above theoretical predictions prior to \(\langle E_n \rangle\) attaining \(E^*\). This global offset originates from slight numerical deviations from identical expected stored energy among the four conditional post-measurement battery states. Individual trajectories nevertheless remain nearly identical to one another, yielding \(\sigma(E_n) < 5.0\times 10^{-14}\), consistent with vanishing fluctuations within numerical precision.

\begin{figure}[!htbp]
\centering
 \includegraphics[width=0.48\columnwidth]{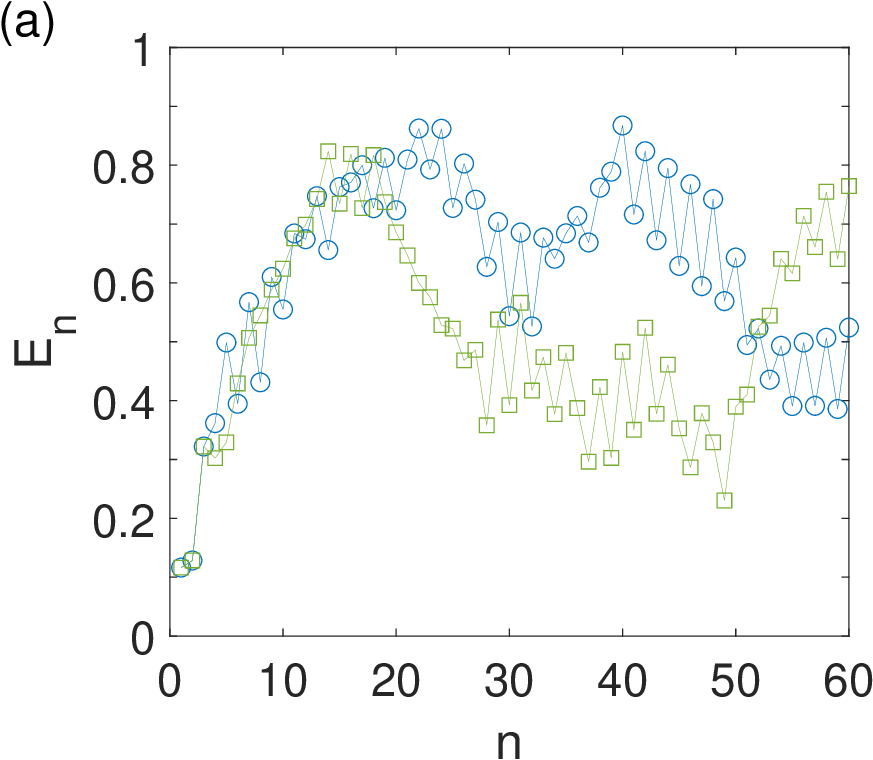}
 \includegraphics[width=0.48\columnwidth]{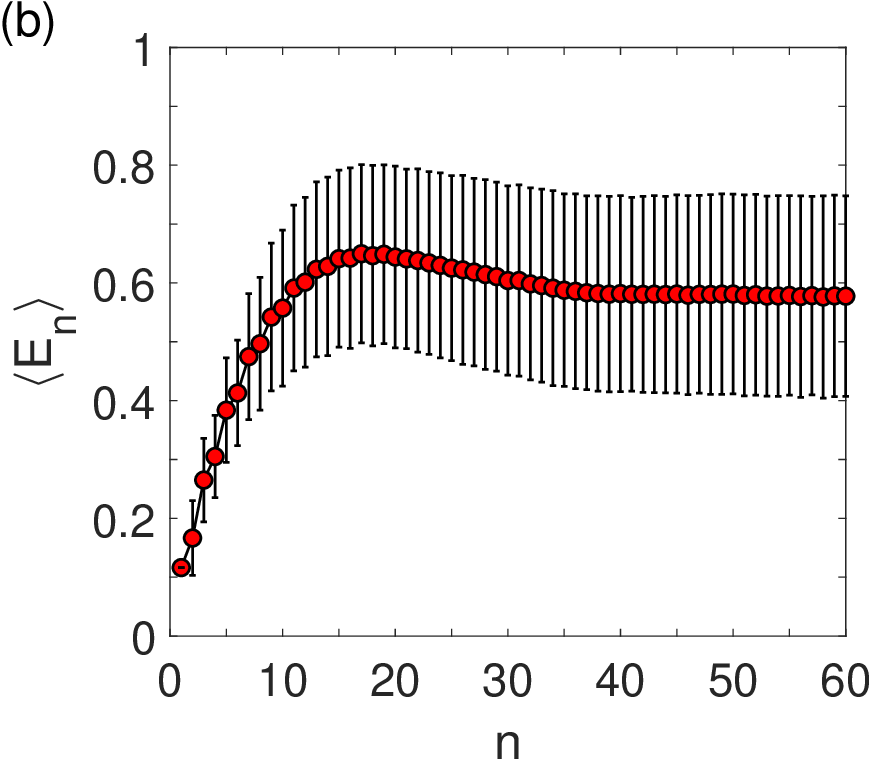}
\caption{(a) Two trajectories of stored energy \(E_n\). (b) Trajectory-averaged stored energy \(\langle E_n\rangle\) versus collision steps \(n\), with error bars for stored-energy fluctuations \(\sigma(E_n)\). Parameters: \(\Delta=0.7\) and \(g=0.05\); \(\gamma_{01}=\pi/3\), \(\gamma_{10}=\pi/5\), and \(\gamma_{11}=\pi/7\); \(e_{00}=0.35\), \(e_{01}=0.05\), \(e_{10}=0.10\), and \(e_{11}=0.5\). The corresponding concurrence \(C=0.8001\).} \label{Fig3}
\end{figure}

(iii) Other entangled states. The charger state in Eq.~\eqref{EQ_phia} is entangled when its concurrence \(C>0\). Apart from the two families of Bell-like entangled states, Eq.~\eqref{EQ_en1} shows that \(e_{n+1}\) depends on \(\gamma_n\), the phase of the post-measurement battery state \(|\phi_b(n\tau)\rangle\), see Eq.~\eqref{EQ_phib}. Here, \(\gamma_n\) corresponds to \(\gamma_m\) for the measurement branch \(m\) realized in this collision. Since the four conditional states \(|\phi_b^0\rangle,|\phi_b^1\rangle,|\phi_b^2\rangle,|\phi_b^3\rangle\) may carry different phases \(\gamma_0,\gamma_1,\gamma_2,\gamma_3\) [see Eq.~\eqref{EQ_CSN}], the subsequent descendant branches will generally yield different values for \(e_{n+1}\). By mathematical induction, distinct battery energies generally develop among trajectory branches after the \(n\)-th collision. This implies that stored-energy fluctuations exist. Figure~\ref{Fig3} exemplifies this behavior: Panel (a) shows that two individual energy trajectories remain distinct, directly demonstrating the presence of fluctuations; panel (b) reveals that fluctuations \(\sigma(E_n)\) become pronounced at large collision steps \(n\).

The recurrence relation Eq.~\eqref{EQ_en1} for \(e_n\) involves the trajectory-dependent battery phase \(\gamma_n\), together with the populations and intrinsic relative phases of the charger state. This intricate multiparameter interference structure makes it difficult to establish clear relations between the trajectory-averaged stored energy \(\langle{E_n}\rangle\) and the concurrence \(C\).

\emph{Separable states}---A two-qubit state is separable when its concurrence \(C=0\), see Eq.~\eqref{EQ_CC}. For the charger state in Eq.~\eqref{EQ_phia}, that means
\begin{equation}
\sqrt{e_{00}e_{11}}=\sqrt{e_{01}e_{10}},\quad \gamma_{11}=(\gamma_{01}+\gamma_{10})\bmod 2\pi.
\label{EQ_ss}
\end{equation}

(i) Computational basis states. A two-qubit system has four computational basis states, \(|\phi_a\rangle=|00\rangle\), \(|01\rangle\), \(|10\rangle\), and \(|11\rangle\). These are the separable computational-basis limit of the Bell-like families, and we refer to them as Bell-like separable states. In fact, \(e_n\) can still be calculated using Eq.~\eqref{EQ_Edef_BS1} for the first and fourth states. The first state is a trivial case as \(e_n=0\) for all \(n\); the fourth state is the benchmark example introduced earlier. For the second and third states, \(e_n\) can be calculated using Eq.~\eqref{EQ_Edef_BS2}. These \(e_n\) are independent of the battery phase \(\gamma_n\), so stored-energy fluctuations are completely suppressed.

\begin{figure}[!htbp]
\centering
 \includegraphics[width=0.48\columnwidth]{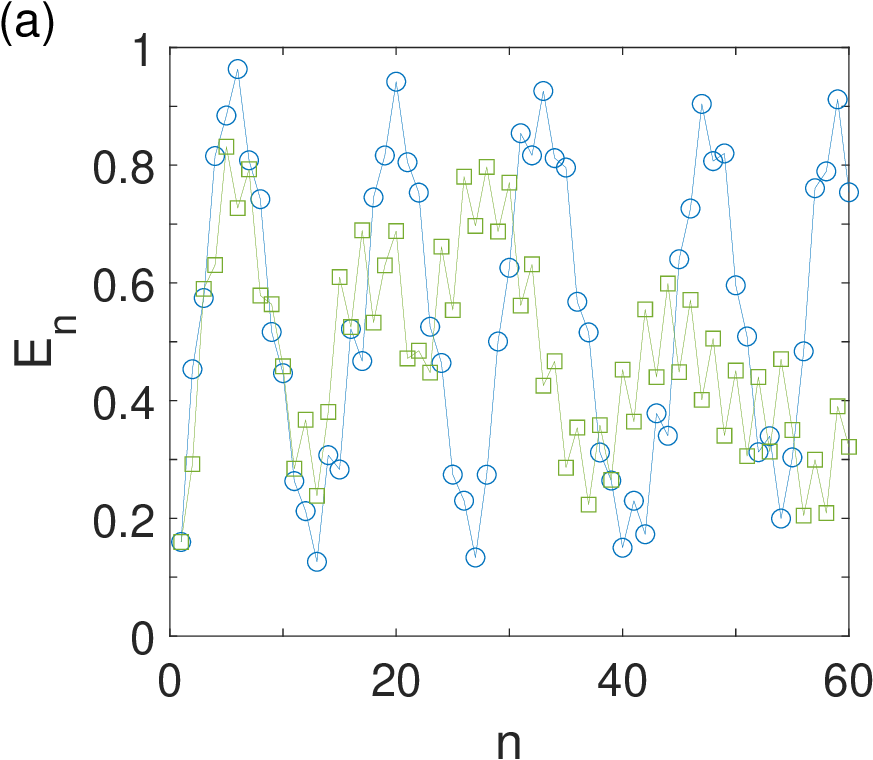}
 \includegraphics[width=0.48\columnwidth]{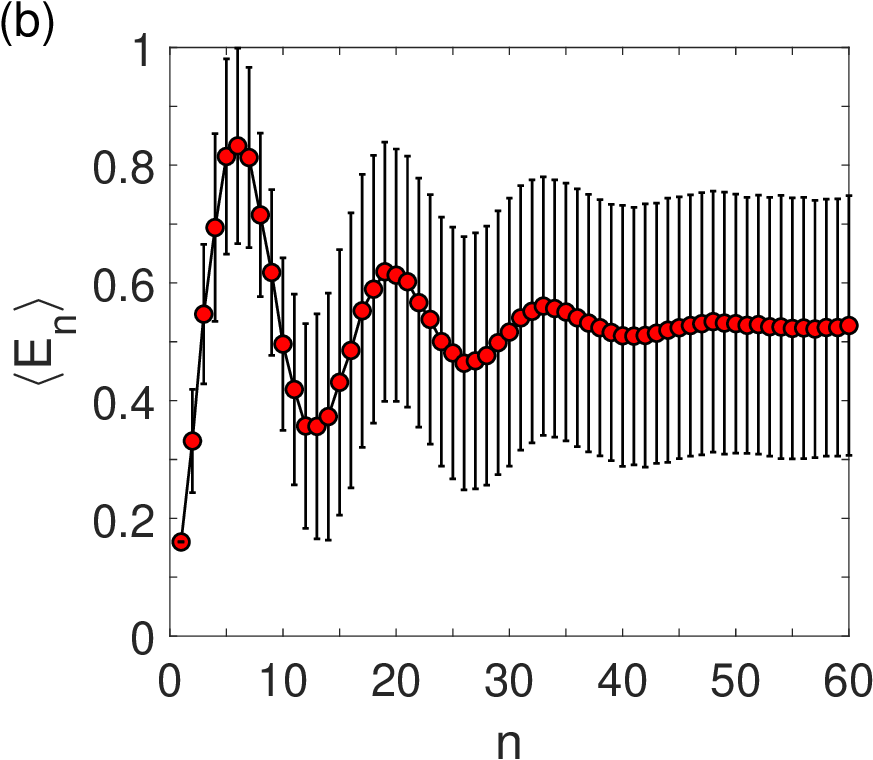}
\caption{(a) Two trajectories of stored energy \(E_n\). (b) Trajectory-averaged stored energy \(\langle E_n\rangle\) versus collision steps \(n\), with error bars for stored-energy fluctuations \(\sigma(E_n)\). Parameters: \(\Delta=0.7\) and \(g=0.05\); \(\gamma_{01}=\pi/3\), \(\gamma_{10}=\pi/5\), and \(\gamma_{11}=8\pi/15\); \(e_{00}=0.08\), \(e_{01}=0.12\), \(e_{10}=0.32\), and \(e_{11}=0.48\). The corresponding concurrence \(C=0\).} \label{Fig4}
\end{figure}

(ii) Superposition states in the computational basis. If a state satisfies Eq.~\eqref{EQ_ss}, then $e_n$ includes multiple cosine terms of $\gamma_n$. While destructive interference is possible, full cancellation seldom happens, so stored-energy fluctuations generally persist. Figure~\ref{Fig4} illustrates an example. Panel (a) shows that the two individual energy trajectories are distinct; panel (b) shows significant fluctuations in \(E_n\) at relatively large collision steps \(n\).

\emph{Conclusions and outlook}---Initial-state entanglement is actively investigated for quantum battery charging. We identify a criterion for stored-energy fluctuations across distinct trajectories in cascaded collisional quantum batteries: Fluctuations appear when the recurrence relation of stored energy depends on the battery phase $\gamma_n$, and vanish otherwise. Fluctuations vanish for chargers drawn from the two Bell-like families of entangled states, as well as Bell-like separable states; fluctuations appear for all remaining charger states, whether the state is entangled or separable. Adaptive post-collision measurements ensure uniform expected stored energy across measurement outcomes, mitigating intrinsic quantum randomness. Future work may explore larger architectures and test their robustness against measurement imperfections and environmental dissipation.

\emph{Acknowledgments}---This work was supported by the National Natural Science
Foundation of China (Grant No. 12674301).

\emph{Data availability}---There are no publicly available
research data or software supporting this manuscript.
Requests for further information or data should be sent
to the authors.

\begin{thebibliography}{99}
\bibitem{CA24}F. Campaioli, S. Gherardini, J. Q. Quach, M. Polini, and G. M. Andolina,
Colloquium: Quantum batteries, Rev. Mod. Phys. {\bf 96},~031001~(2024).
\bibitem{FE26}D. Ferraro, F. Cavaliere, M. G. Genoni, G. Benenti, and M. Sassetti,
Opportunities and challenges of quantum batteries, Nat. Rev. Phys. {\bf 8},~115~(2026). 
\bibitem{AU22}A. Auff\`{e}ves, Quantum technologies need a quantum energy initiative, PRX Quantum {\bf 3},~020101~(2022).
\bibitem{CA25}A. Camposeo, T. Virgili, F. Lombardi, G. Cerullo, D. Pisignano, and M. Polini,
Quantum batteries: A materials science perspective, Adv. Mater. {\bf 37},~2415073~(2025). 
\bibitem{JO22}J. Joshi, T. S. Mahesh, Experimental investigation of a quantum battery using star-topology NMR spin systems, Phys. Rev. A {\bf 106},~042601~(2022). 
\bibitem{YU24}J. Yu, S. Wang, K Liu, et al., Experimental demonstration of a Maxwell's demon quantum battery in a superconducting noisy intermediate-scale quantum processor, Phys. Rev. A {\bf 109},~062614~(2024). 
\bibitem{ZH23}G. Zhu, Y. Chen, Y. Hasegawa, and P. Xue, Charging quantum batteries via indefinite causal order: Theory and experiment, Phys. Rev. Lett. {\bf 131},~240401~(2023).
\bibitem{LU25}Z.-G. Lu, G. Tian, X.-Y. L\"{u}, and C. Shang, Topological quantum batteries, Phys. Rev. Lett. {\bf 134},~180401~(2025).
\bibitem{QU22}J. Q. Quach, K. E. McGhee, L. Ganzer, D. M. Rouse, B. W. Lovett, E. M. Gauger, J. Keeling, G. Cerullo, D. G. Lidzey, and T. Virgili, Superabsorption in an organic microcavity: Toward a quantum battery, Sci. Adv. {\bf 8},~eabk3160~(2022). 
\bibitem{HY26}K. Hymas, J. B. Muir, D. Tibben, J. van Embden, T. Hirai, C. J. Dunn, D. E. G\'{o}mez, J. A. Hutchison, T. A. Smith, and J. Q. Quach, Superextensive electrical power from a quantum battery, Light: Sci. Appl. {\bf 15},~168~(2026). 
\bibitem{TA26}N. Ta\c{s}alt{\i}n and D. E. Y{\i}ld{\i}z, Quantum batteries: Energy storage models and physical approaches via quantum systems, J. Energy Storage {\bf 162},~122088~(2026). 
\bibitem{KH26}B. Khadim, A. Majid, A. Belgibayeva, Y. Jin, N. Bulut, and M. Alkhedher, Quantum batteries: Unlocking the future of high-tech energy storage, J. Energy Storage {\bf 141},~119249~(2026). 
\bibitem{AL13}R. Alicki and M. Fannes, Entanglement boost for extractable work from ensembles of quantum batteries, Phys. Rev. E {\bf 87},~042123~(2013). 
\bibitem{SH25}A. Shokri, E. Faizi, and M. B. Arjmandi, Entanglement-assisted charging of quantum batteries within optomechanical framework, Phys. Rev. E {\bf 111},~064117~(2025).
\bibitem{KA20}F. H. Kamin, F. T. Tabesh, S. Salimi, and A. C. Santos, Entanglement, coherence, and charging process of quantum batteries, Phys. Rev. E {\bf 102}, 052109 (2020).
\bibitem{LI21}J. X. Liu, H. L. Shi, Y. H. Shi, X. H. Wang, and W. L. Yang, Entanglement and work extraction in the central-spin quantum battery, Phys. Rev. B {\bf 104},~245418~(2021).
\bibitem{SH26}R. K. Shukla, S. K. Mishra, and U. Sen, Collective dynamics versus entanglement in quantum battery performance, Phys. Rev. A {\bf 114},~022624~(2026).
\bibitem{AR22}M. B. Arjmandi, A. Shokri, E. Faizi, and H. Mohammadi, Performance of quantum batteries with correlated and uncorrelated chargers, Phys. Rev. A {\bf 106},~062609~(2022).
\bibitem{MA24}H.-B. Ma, K. Xu, H.-G. Li, Z.-G. Li, and H.-J. Zhu, Enhancing the charging performance of quantum batteries with the work medium of an entangled coupled-cavity array, Phys. Rev. A {\bf 110},~022433~(2024).
\bibitem{GY24}J.-Y. Gyhm and U. R. Fischer, Beneficial and detrimental entanglement for quantum battery charging, AVS Quantum Sci. {\bf 6},~012001~(2024).
\bibitem{LI26} H. Li, X. Chen, W. Yu, and Y. Hao, Charging optimization of quantum battery via phase-controlled coherent ancillae, Phys. Rev. A {\bf 113},~052437~(2026).
\bibitem{FH18}N. Friis and M. Huber, Precision and work fluctuations in Gaussian battery charging, Quantum {\bf 2},~61~(2018).
\bibitem{CR20}A. Crescente, M. Carrega, M. Sassetti and D. Ferraro, Charging and energy fluctuations of a driven quantum battery, New J. Phys. {\bf 22},~063057~(2020).
\bibitem{MI21}M. T. Mitchison, J. Goold and J. Prior, Charging a quantum battery with linear feedback control, Quantum {\bf 5},~500~(2021).
\bibitem{GO26}L. Gong, J. Zhang, and Y. Li, Achieving identical stored energy in cascaded collisional quantum battery charging: Analytical results, arXiv preprint arXiv:2610.01332
     (2026).
\bibitem{CI22}F. Ciccarello, S. Lorenzo, V. Giovannetti, and G. M. Palma, Quantum collision models: Open system dynamics from repeated interactions, Phys. Rep. {\bf 954},~1~(2022).
\bibitem{HI97}S. A. Hill and W. K. Wootters, Entanglement of a pair of quantum bits, Phys. Rev. Lett. {\bf 78},~5022~(1997).
\bibitem{WO98}W. K. Wootters, Entanglement of formation of an arbitrary state of two qubits, Phys. Rev. Lett. {\bf 80},~2245~(1998).

\end{thebibliography}


\end{document}